\documentclass[preprint]{revtex4-2}
\usepackage[dvips]{graphicx}
\usepackage{color}
\usepackage{amsmath}
\usepackage{amssymb}
\usepackage{bm}
\usepackage{textcomp}

\begin{document}

\title{Fluctuating Diffusivity Emerges even in Binary Gas Mixtures}

\author{Fumiaki Nakai}
\email{nakai.fumiaki.c7@s.mail.nagoya-u.ac.jp}
\author{Yuichi Masubuchi}
\author{Yuya Doi}
\author{Takato Ishida}
\author{Takashi Uneyama}
\email{uneyama@mp.pse.nagoya-u.ac.jp}
\affiliation{Department of Materials Physics, Graduate School of Engineering, Nagoya University,
Furo-cho, Chikusa, Nagoya 464-8603, Japan}

\begin{abstract}
Diffusivity in some soft matter and biological systems changes with time, called the fluctuating diffusivity.
In this work, we propose a novel origin for fluctuating diffusivity based on stochastic simulations of binary gas mixtures.
In this system, the fraction of one component is significantly small, and the mass of the minor component molecule is different from that of the major component.
The minor component exhibits fluctuating diffusivity when its mass is sufficiently smaller than that of the major component. We elucidate that this fluctuating diffusivity is caused by the time scale separation between the relaxation of the velocity direction and the speed of the minor component molecule.

\end{abstract}
\maketitle
\newpage

\section{Introduction}
Brownian motion is widely observed in soft matter systems, and standard Brownian motion is described by a simple stochastic process known as the Wiener process\cite{vanKampen2007, gardiner2009stochastic}. In this process, the mean square displacement (MSD) increases linearly with time and is
accompanied by a Gaussian displacement distribution\cite{zwanzig2001nonequilibrium, nelson2020dynamical}.
Although this simple Brownian motion is fully understood, a new type of Brownian motion has been recently reported; although the MSD is proportional to time, the displacement distribution deviates from the Gaussian distribution\cite{wang2009anomalous, he2016dynamic, guan2014even, uneyama2015fluctuation}. This motion is known as Brownian (or Fickian) yet non-Gaussian diffusion and cannot be described by the simple Wiener process.
This process can be successfully
described by the Langevin equation with the time-dependent fluctuating diffusivity\cite{uneyama2015fluctuation}.
\begin{equation}
 \label{lefd}
 \frac{d\bm{R}(t)}{dt}=\sqrt{2D(t)}\bm{\xi}(t),
\end{equation}
where $\bm{R}(t)$ denotes the position of the Brownian particle, $D(t)$ denotes the fluctuating diffusivity, and
$\bm{\xi}(t)$ is Gaussian white noise. 
The fluctuating diffusivity obeys a stochastic process independent of $\bm{R}(t)$.
The first and second order statistical moments of $\bm{\xi}(t)$ are given as $\langle \bm{\xi}(t) \rangle = \bm{0}$ and $\langle \bm{\xi}(t) \bm{\xi}(t') \rangle = \bm{I} \delta(t - t')$, where $\langle \dots \rangle$ represents the statistical average and $\bm{I}$ is the unit tensor.

The origins of the fluctuating diffusivity in soft matter and biological systems can be classified into two categories\cite{uneyama2019relaxation}.
The first origin is a spatially and/or temporally heterogeneous environment\cite{chechkin2017brownian, chubynsky2014diffusing}.
For instance, particles in supercooled liquids (glass formers)\cite{kob1997dynamical, yamamoto1998heterogeneous, miyaguchi2016langevin}, colloidal suspensions \cite{guan2014even, kim2013simulation, pastore2021rapid}, biological systems \cite{he2016dynamic, wang2009anomalous, jeon2016protein, rusciano2022fickian}, and active matter \cite{leptos2009dynamics, kurtuldu2011enhancement} exhibit fluctuating diffusivities, owing to their heterogeneous environments.
The second origin is the fluctuation in the conformational degrees of freedom.
That is, the diffusivity can fluctuate depending on the fluctuations of the conformation or orientation of a molecule\cite{yamamoto2021universal, uneyama2015fluctuation, miyaguchi2017elucidating}. Examples include the center of mass of an entangled polymer\cite{uneyama2015fluctuation} and rod-like particle solution\cite{miyaguchi2017elucidating}.

Here, one question may arise: are there only two origins of fluctuating diffusivity?
In this study, we demonstrate that the third origin of fluctuating diffusivity exists by investigating simple gas systems, i.e., binary gas mixtures comprising hard spheres with different masses, in which the fraction of one component is sufficiently small.
These systems do not possess a heterogeneous environment nor conformational degrees of freedom, which are known to be the origins of fluctuating diffusivity.
The gas molecules are assumed to be spherical and do not have any internal degrees of freedom. They are randomly distributed in space, and there is no spatial correlation.
Even in such systems, the fluctuating diffusivity causing Brownian yet non-Gaussian diffusion emerges under specific conditions.
We elucidate that the observed fluctuating diffusivity originates from the separation of time scales of two relaxation processes of the minor component; the velocity direction relaxation and speed relaxation.

\section{System}
The dynamics of a single molecule $A$ in another gas molecule $B$ is investigated as a model of binary gas mixtures, where the fraction of molecules of gas $A$ is sufficiently small. The molecules $A$ and $B$ have different masses, $m_A$ and $m_B$, and sizes $\sigma_A$ and $\sigma_B$, respectively.
The system is in equilibrium with inverse temperature $\beta$, and the number density of molecule $B$ is $\rho$.
Molecule $A$ moves ballistically until it collides with molecule $B$. Molecule $A$ instantaneously changes its velocity by collision based on the conventional hard-sphere interaction \cite{dorfman2021, allen2017} as follows:
\begin{equation}
 \bm{v}'_A = \bm{v}_A-\frac{2m_B}{m_B+m_A}\left(\bm{v}_A-\bm{v}_B\right)\cdot \hat{\bm{r}}_{AB}\hat{\bm{r}}_{AB}.
\label{eq_velocity_change_tracer}
\end{equation}
Here, $\bm{v}_{A}'$ is the velocity of molecule $A$ after collision, $\bm{v}_{A}$ and $\bm{v}_{B}$ are the velocities of molecules $A$ and $B$ before collision, respectively, and $\hat{\bm{r}}_{AB}$ is the unit vector connecting the centers of molecules $A$ and $B$.
Here, it should be mentioned that this collision protocol is not crucial for the following results; similar data will be obtained for other interaction potentials such as the Weeks-Chandler-Andersen potential.

In gas systems, the dynamics of a molecule can be approximately described as a Markovian stochastic process because the dynamic correlations are weak\cite{ehrenfest1990conceptual, dorfman2021, chapman1990}.
Therefore, we employ the kinetic Monte Carlo (KMC) method \cite{gillespie1976general, bortz1975new} to simulate the dynamics of molecule $A$.
Collision statistics are required for implementing the KMC method. 
In hard-sphere gas, the probability density of molecule $A$ colliding with molecule $B$ with $\bm{v}_B$ at $\hat{\bm{r}}$ and time interval $s$ for a given $\bm{v}_A$ becomes
\begin{equation}
\begin{split}
 &P(\bm{v}_B, \hat{\bm{r}}_{AB}, s|\bm{v}_A)
 \\
= & \rho\sigma^2(\bm{v}_B-\bm{v}_A)\cdot\hat{\bm{r}}_{AB}
 \left(\frac{\beta  m_{B}}{2 \pi}\right)^{3/2}
 \exp\left(-\frac{\beta  m_{B} \bm{v}_{B}^{2}}{2}\right)
 \\
& \times\exp[-F( \bm{v}_A) s]  \Theta[(\bm{v}_A-\bm{v}_B)\cdot \hat{\bm{r}}_{AB}].
\end{split}
\label{eq_collision_probability}
\end{equation}
Here, $\sigma = (\sigma_A+\sigma_B)/2$, $F( \bm{v}_A)$ is the average collision frequency of molecule $A$ with velocity $\bm{v}_{A}$, and $\Theta(x)$ is the Heaviside step function (collision does not occur for $(\bm{v}_A-\bm{v}_B)\cdot \hat{\bm{r}}_{AB}< 0$).
Here, we emphasize that Eq.~\eqref{eq_collision_probability} does not depend on the spatial position nor time; the statistics depend only on the velocity of molecule A.
The explicit expression of $F( \bm{v}_A)$, derivation of Eq.~\eqref{eq_collision_probability}, and numerical scheme are explained in Appendix~\ref{appendix_collision_statistics} and \ref{appendix_scheme}.
The dynamics of molecule $A$ can be characterized only by the mass ratio $\mu=m_{A} / m_{B}$.
We employ dimensionless units by setting $m_B = 1$, $\beta^{-1} = 1$, and $1/\rho\sigma^2 = 1$.

\section{Numerical results}
\begin{figure}[htbp]	
\begin{center}
  \includegraphics[width=0.5\linewidth]{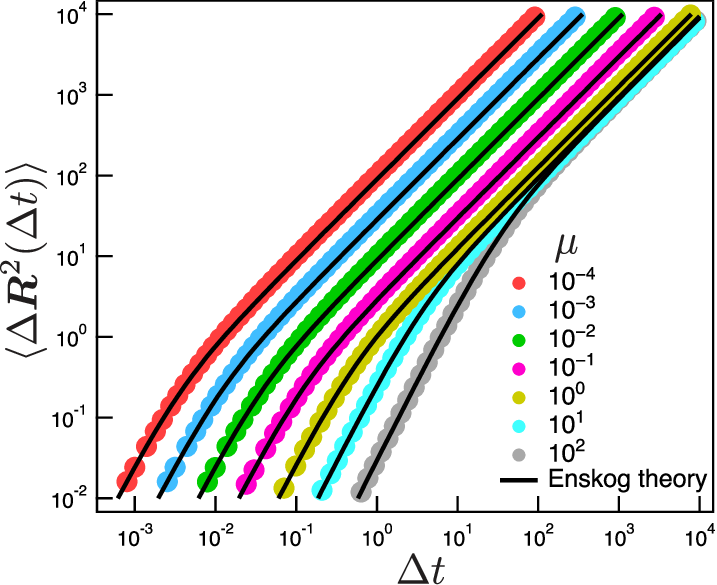}
\caption{Mean square displacements (MSDs) of the molecule $A$ for several mass ratios $\mu$. The symbols are the KMC simulation data, and the black solid curves represent the prediction by the Enskog theory
(Eq.~\eqref{eq_enskog_msd}).}
\label{fig_msd}
\end{center}
\end{figure}

Figure~\ref{fig_msd} shows the MSD $\langle \Delta\bm{R}^{2}(\Delta t) \rangle$, where $\Delta \bm{R}(\Delta t) = \bm{R}(\Delta t) - \bm{R}(0)$ and $\Delta t$ denotes the time lag.
For comparison, we have included the prediction by the Enskog theory\cite{alder1974studies, chapman1990}:
\begin{equation}
 \langle \Delta \bm{R}^2(\Delta t) \rangle
=\frac{3\tau_c^2}{2\mu}\left[-1+ \frac{2\Delta t}{\tau_c}
  +e^{-{2\Delta t}/{\tau_c}}\right],\label{eq_enskog_msd}
\end{equation}
where $\tau_{c}$ is the crossover time from ballistic to diffusive
regions defined as follows:
\begin{equation}
 \tau_{c} =\sqrt{9\mu(\mu+1) / 32 \pi }.\label{eq_tau_c}
\end{equation}
The results obtained from the KMC simulations exhibit simple ballistic and diffusive behaviors in the simulated $\mu$ range, and these results are almost perfectly reproduced by the Enskog theory.
We naively expect that the dynamics of molecule $A$ is simple Brownian motion with constant diffusivity for any $\mu$.

\begin{figure}[htbp]
 \begin{center}
  \includegraphics[width=0.5\linewidth]{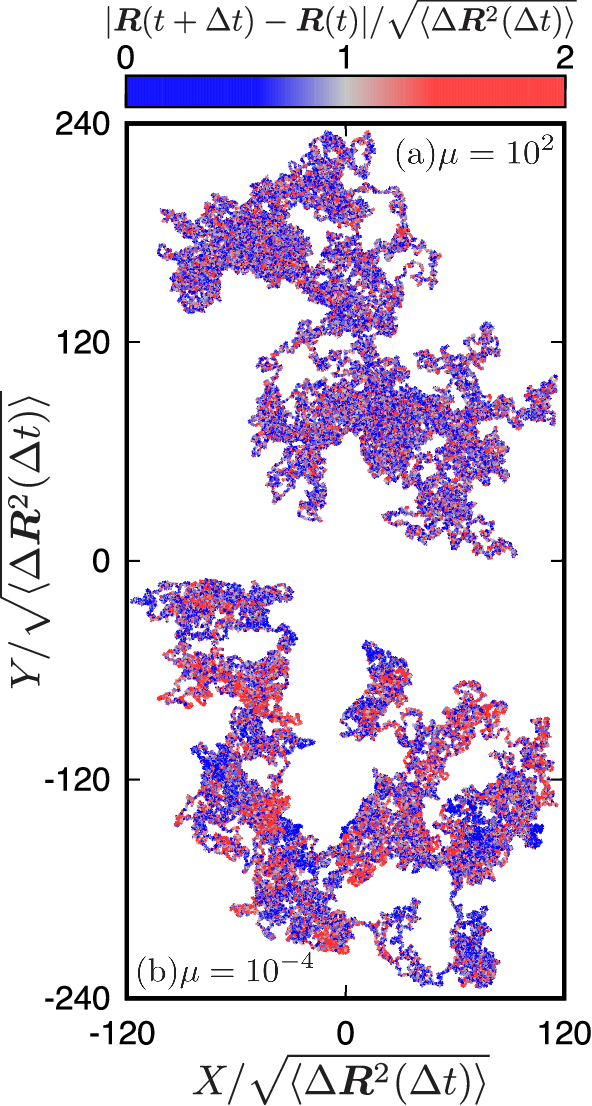}
\caption{
Typical trajectories of the molecule $A$ during $0 \le t \le 10^{6} \tau_{c}$
for (a)$\mu = 10^{2}$ and (b)$10^{-4}$ from the KMC simulation.
The trajectories are mapped
onto the $xy$ plane.
The colors represent the reduced temporal displacement
$|\bm{R}(t + \Delta t) - \bm{R}(t)| / \sqrt{\langle \Delta \bm{R}^2(\Delta t) \rangle}$ with $\Delta t = 10 \tau_{c}$.}
\label{fig_trajectory}
 \end{center}
\end{figure}

However, the dynamics of molecule $A$ is not simple Brownian motion for small $\mu$.
Figure~\ref{fig_trajectory} shows the trajectories of molecule $A$ for sufficiently large and small mass ratios $\mu=10^2$ and $10^{-4}$.
The observation time is $T=10^6\tau_c$, and the trajectories are mapped onto the $xy$ plane.
The colors express the magnitude of the scaled temporal displacement for a time lag $\Delta t=10\tau_c$.
For $\mu=10^2$, the fast (red) and slow (blue) areas are homogeneously distributed; this is consistent with simple Brownian motion.
By contrast, for $\mu=10^{-4}$, large clusters of fast and slow areas are clearly observed.
This implies that the dynamics of molecule $A$ deviates from a simple Brownian motion when $\mu$ is small.
In what follows, we present the results with typical mass ratios, $\mu=10^2$ and $10^{-4}$, as the representative cases of simple Brownian motion and non-trivial diffusion, respectively.
Data for other mass ratios are summarized in Appendix~\ref{appendix_data}.

\begin{figure}[htbp]
 \begin{center}
\includegraphics[width=0.5\linewidth]{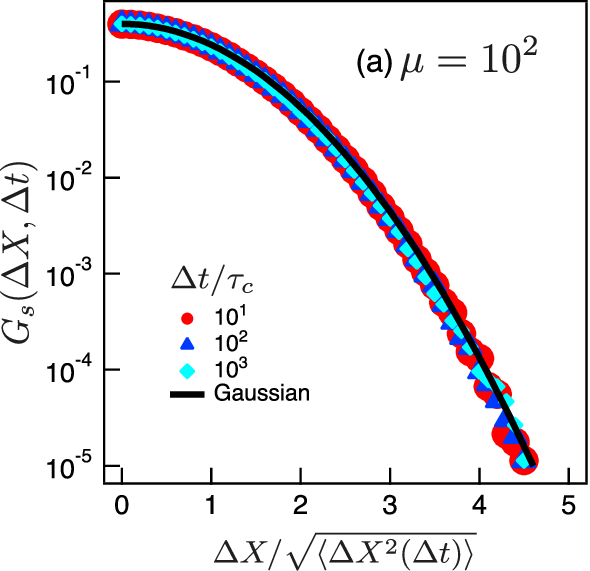}
  \includegraphics[width=0.5\linewidth]{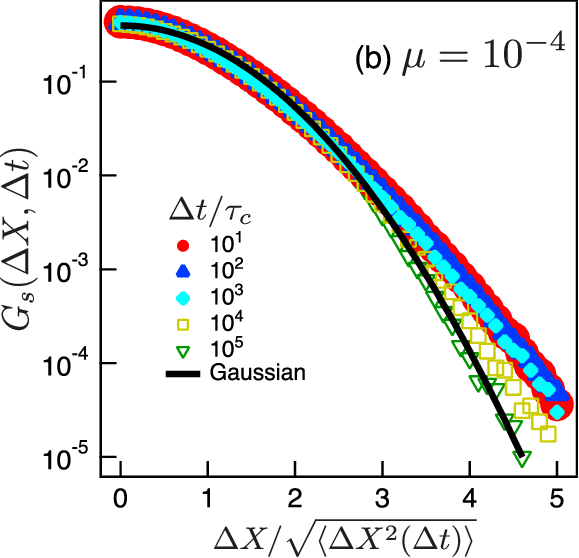}
\caption{Self-part of the van Hove correlation functions of the molecule $A$ for different time lags $\Delta t$ from the KMC simulation.
  (a) $\mu = 10^{2}$ and (b) $10^{-4}$.
  For convenience, the displacement is normalized using the root MSD $\sqrt{\langle \Delta X^{2}(\Delta t) \rangle}$. The solid black curves represent the Gaussian distribution.
  }
\label{van_hove}
 \end{center}
\end{figure}

To examine whether the dynamics of molecule $A$ is Gaussian, we calculate the self-part of the van Hove correlation functions,
which is defined as $G_{s}(\Delta X,\Delta t)= \langle \delta[\Delta X - (X(t+\Delta t) - X(t))]\rangle$, where $X(t)$ is the position of molecule $A$ in the $x$ direction at time $t$.
Figure~\ref{van_hove} shows $G_{s}(\Delta X,\Delta t)$ for various $\Delta t$.
For $\mu=10^2$, $G_{s}(\Delta X,\Delta t)$ is Gaussian within the simulated $\Delta t$ range. In contrast, for $\mu = 10^{-4}$, $G_{s}(\Delta X,\Delta t)$ deviates from the Gaussian distribution within an intermediate time lag, $10^{1} \lesssim \Delta t/\tau_c \lesssim 10^{4}$. This deviation disappears for a sufficiently large time lag $\Delta t/\tau_c \gtrsim 10^5$.
Therefore, Brownian yet non-Gaussian diffusion appears for $\mu=10^{-4}$ at the intermediate time scale.
This behavior is commonly observed for $\mu<1$ as shown in Fig.~\ref{fig_appendix_van_hove} in Appendix.
The non-Gaussian behavior can be also observed in the non-Gaussian parameter (NGP) shown in Fig.~\ref{fig_appendix_ngp} in Appendix.
The NGP exhibits non-negligible peaks for $\mu < 1$.

\begin{figure}[htbp]
\begin{center}
 \includegraphics[width=0.5\linewidth]{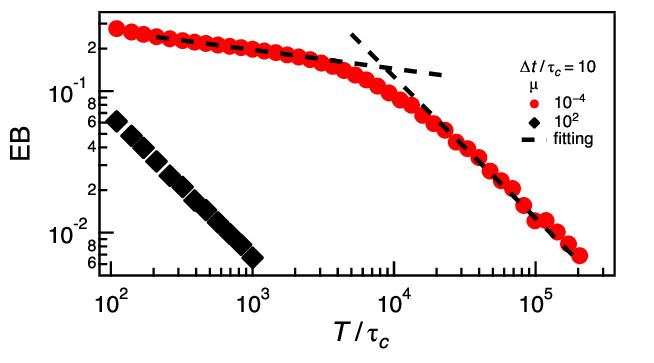}
 \caption{
Ergodicity breaking (EB) parameters corresponding to $\mu = 10^{2}$ and $10^{-4}$ from the KMC simulation. The time lag is set as $\Delta t = 10 \tau_{c}$.
The dotted lines indicate the fitting results to the power-laws $\text{EB} \propto T^{-\alpha}$ and $\text{EB} \propto T^{-1}$.
\label{fig_eb}}
\end{center}
\end{figure}

To analyze the non-Gaussian behavior in detail, we calculate the ergodicity breaking (EB) parameter\cite{uneyama2015fluctuation, cherstvy2013anomalous} defined as follows:
\begin{equation}
 \mathrm{EB}(\Delta t, T)=\frac{\Big\langle \Big[\overline{\delta^2}(\Delta t, T)\Big]^{2}\Big\rangle}{\Big\langle \overline{\delta^2}(\Delta t, T) \Big\rangle^2}-1. \label{eq_eb}
\end{equation}
Here, $\overline{\delta^2}(\Delta t, T)$ denotes the time-averaged MSD
for the time lag $\Delta t$ and finite observation time $T$:
\begin{equation}
 \overline{\delta^2}(\Delta t, T) = \frac{1}{T-\Delta t}\int^{T-\Delta t}_0[\bm{R}(t+\Delta t)-\bm{R}(t)]^2 dt.
\end{equation}
The dependence of the EB parameter on $\Delta t$ was theoretically 
proven to be weak when $T\gg \Delta t$ \cite{uneyama2015fluctuation}. Therefore, we set $\Delta t/\tau_c=10$ and calculate the EB parameter as a function of $T$ for $T/\tau_c\ge 10^2$.
Figure~\ref{fig_eb} displays the observation time dependence of the EB parameter, which simply exhibits a decay $\text{EB} \propto T^{-1}$ in the entire $T$ range for $\mu = 10^{2}$. This implies that the dynamics of molecule $A$ follows a Gaussian process. In contrast, for $\mu=10^{-4}$, the EB parameter exhibits a shoulder before the Gaussian decay $\text{EB} \propto T^{-1}$.
This is also observed for other sufficiently small mass ratios, $\mu\ll 1$, as shown in Fig.~\ref{fig_appendix_eb} in Appendix.
The existence of this shoulder can be attributed to the fluctuating diffusivity \cite{uneyama2015fluctuation}, and the characteristic crossover time $\tau_{\text{EB}}$ from the shoulder to the $\text{EB} \propto T^{-1}$ decay can be interpreted as the relaxation time of the fluctuating diffusivity\cite{uneyama2015fluctuation}.
The crossover time $\tau_{\text{EB}}$ for $\mu=10^{-4}$ is estimated
from the two curve fittings $\text{EB} \propto T^{-\alpha}$
where $0 < \alpha < 1$ for short $T$ and $\text{EB} \propto T^{-1}$ for long $T$ regions. The obtained $\tau_{\text{EB}}$ for $\mu=10^{-4}$ is approximately equal to the time scale at which the van Hove correlation function becomes Gaussian.

\section{Origin of the fluctuating diffusivity}
\begin{figure}[htbp]
 \begin{center}
  \includegraphics[width=0.5\linewidth]{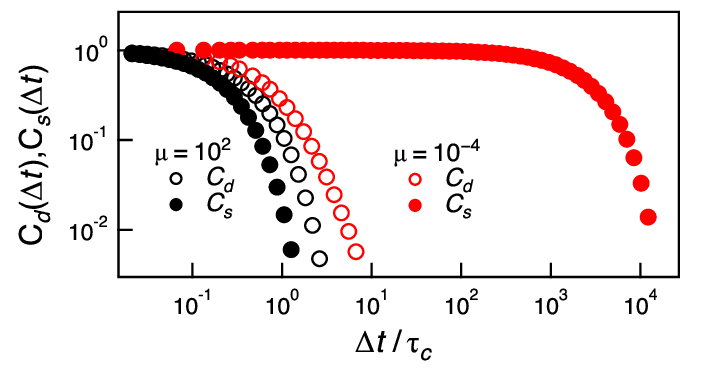}
\caption{Correlation functions of the velocity direction $C_d(\Delta t)$ and speed $C_s(\Delta t)$ of molecule $A$ (Eqs.~\eqref{eq_dcf} and \eqref{eq_scf}) for $\mu=10^{-4}$ and $10^2$ from the KMC simulation.}
\label{fig_direction_speed}
 \end{center}
\end{figure}

When $\mu$ is sufficiently small, i.e., $\mu\ll 1$, the velocity of molecule $A$ is significantly larger than that of molecule $B$, i.e., $|\bm{v}_A|\gg |\bm{v}_B|$.
Under such a condition, the motion of molecule $A$ is similar to that in a matrix of immobile obstacles such as Lorentz gases\cite{andersen1964relaxation, boldrighini1983boltzmann, dorfman2021, moran1987diffusion, machta1983diffusion}.
The speed of molecule $A$ is nearly unchanged by a few collisions, whereas the velocity direction is randomized.
Therefore, we expect that the relaxation times of the speed and velocity direction of molecule $A$ will be considerably different if $\mu$ is small.
We calculate the correlation functions corresponding to the velocity direction $C_{d}(\Delta t)$ and speed $C_{s}(\Delta t)$:
\begin{align}
 C_d(\Delta t) &= \left\langle\frac{\bm{V}(\Delta t)}{|\bm{V}(\Delta t)|}\cdot\frac{\bm{V}(0)}{|\bm{V}(0)|}\right\rangle\label{eq_dcf},\\
C_s(\Delta t) &= \frac{\langle|\bm{V}(\Delta t)||\bm{V}(0)|\rangle - \langle |\bm{V}| \rangle^2}{\langle |\bm{V}|^{2} \rangle - \langle |\bm{V}| \rangle^2}. \label{eq_scf}
\end{align}
Figure~\ref{fig_direction_speed} displays $C_{d}(\Delta t)$ and $C_{s}(\Delta t)$ obtained from the KMC simulations.
The figure clearly reveals that the relaxation of $C_s(\Delta t)$ (filled red symbols) is significantly slower than that of $C_d(\Delta t)$ for $\mu=10^{-4}$ (open red symbols).
This behavior is commonly observed if $\mu$ is sufficiently small as shown in Figs.~\ref{fig_appendix_cd} and \ref{fig_appendix_cs}.
The relaxation times of the direction $\tau_d$ and speed $\tau_s$ can be estimated
from $C_{d}(\Delta t)$ and $C_{s}(\Delta t)$, respectively.
The estimates scaled by $\tau_c$ (Eq.~\eqref{eq_tau_c}) are summarized in Fig.~\ref{fig_appendix_tau}.
For $\mu = 10^{-4}$, $\tau_{d}$ is found to be comparable
to $\tau_{c}$, whereas $\tau_{s}$ is much longer than $\tau_{c}$.
In addition, $\tau_{s}$ is of the same order as $\tau_{\text{EB}}$, which strongly implies that the relaxation of the fluctuating diffusivity in the binary gas mixtures is related to that of the speed of the molecule $A$.
Here, it should be emphasized that such a timescale separation between the velocity direction and speed is not present without ballistic motion.
Thus, the mechanism of the fluctuating diffusivity observed for purely diffusive motions in some heterogeneous environments\cite{wang2009anomalous, guan2014even, jeon2016protein, kim2022fractal} is different from that in our system.

Based on the above results, we propose a possible scenario for the emergence of fluctuating diffusivity in our binary gas mixture with $\mu\ll 1$. At the intermediate time scale $\tau_{d} \lesssim T \lesssim \tau_{s}$, molecule $A$ diffuses because its velocity direction changes randomly. The speed of molecule $A$ remains approximately constant, $|\bm{v}_A(t)| \approx v_{A}$, and thus the diffusion coefficient can be described by a function of constant as $D(t) = D(v_{A})$.
At the long timescale $T \gtrsim \tau_s$, $D(t)$ starts to fluctuate temporarily owing to the fluctuations of $|\bm{v}_{A}(t)|$.
At the very long time scale $T \gg \tau_{s}$, the fluctuation of the diffusivity is smeared out and the Gaussian normal diffusion with the effective diffusion coefficient $D_{\text{eff}} = \langle D \rangle$ is observed.
Therefore, the origin of the fluctuating diffusivity in our system is
the separation of the relaxation timescales of the velocity direction and speed.
This scenario also explains the clusters observed in Fig.~\ref{fig_trajectory}; they reflect the persistence of the molecule A speed within the timescale $\tau_{s}$.

\begin{figure}
    \centering
    \includegraphics[width=0.5\linewidth]{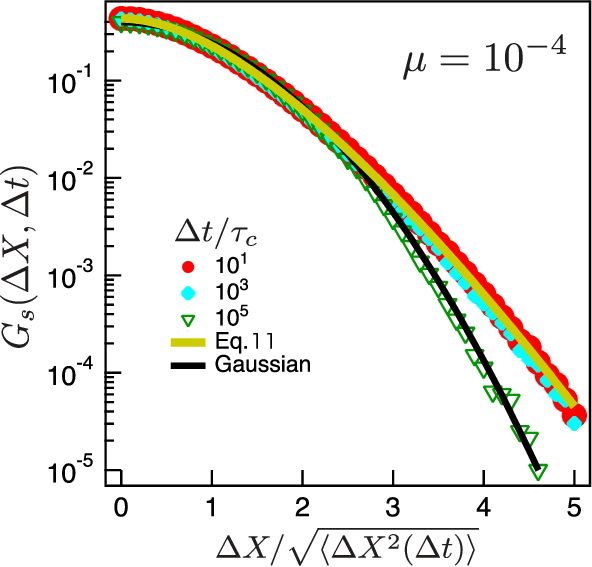}
    \caption{Theoretical prediction of the scaled self-part of the van-Hove correlation function $G_s(\Delta X, \Delta t)$ of the molecule $A$ (ochre curve). For comparison, the KMC simulation results with $\mu=10^{-4}$ for different time lags (Fig.~\ref{van_hove}(b)) are shown with symbols and the Gaussian distribution is displayed with the black curve.}
    \label{fig_van_hove_theory}
\end{figure}

To validate the proposed scenario, we theoretically calculate the van-Hove correlation function of the molecule $A$ with $\mu\ll 1$.
At the intermediate timescale $\tau_{d} \lesssim T \lesssim \tau_{s}$, the dynamics of the molecule $A$ can be virtually described as a mobile particle in dilute fixed spherical obstacles.
Then the diffusion coefficient is calculated as $D(|\bm{v}_A|)= |\bm{v}_A|/3\pi$ \cite{dorfman2021}. The probability density of the displacement of the molecule $A$ under a given speed $v_{A} = |\bm{v}_{A}|$ is Gaussian:
\begin{equation}
    P(\Delta X;\Delta t| v_{A})=\frac{1}{\sqrt{4\pi D(v_{A})
    \Delta t}}\exp\left(-\frac{\Delta X^2}{4 D(v_{A})\Delta t}\right).
    \label{eq_prob_position_given_v}
\end{equation}
In equilibrium, $v_{A}$ obeys the Maxwell-Boltzmann distribution:
$P_{\text{MB}}(v_{A})=4\pi v_{A}^2 (2\pi)^{-3/2} \exp(-v_{A}^{2} / 2)$.
By taking the equilibrium average of Eq.~\eqref{eq_prob_position_given_v} with respect to $v_{A}$, we have
the van-Hove correlation function $G_{s}(\Delta X,\Delta t)$ at the intermediate timescale $\tau_{d} \lesssim \Delta t \lesssim \tau_{s}$:
\begin{equation}
    G_s(\Delta X; \Delta t) = \int_0^{\infty} dv_A P(\Delta X; \Delta t| v_{A})P_{\text{MB}}(v_{A}).
    \label{eq_van_hove_theory}
\end{equation}
We numerically calculate Eq.~\eqref{eq_van_hove_theory} and show the result in 
Fig.~\ref{fig_van_hove_theory}. 
The theoretical prediction by Eq.~\eqref{eq_van_hove_theory} reasonably agrees with the KMC simulation result.
This result supports our scenario on the
fluctuating diffusivity; the fluctuating diffusivity in our system originates from the separation of the relaxation timescales between the velocity direction and the speed.
The tail of $G_{s}(\Delta X;\Delta t)$ from the Gaussian distribution has been
observed in several systems. The tail in Eq.~\eqref{eq_van_hove_theory} can be approximately calculated using the saddle point method:
\begin{equation}
 G_{s}(\Delta X;\Delta t)=
    \sqrt{\frac{3}{4\pi}}\frac{|\Delta X|}{\Delta t}
    \exp\left[-3
	 \left(\frac{3\Delta X^2}{8\sqrt{2}\Delta t}\right)^{\frac{2}{3}}\right] \qquad (\text{for $\Delta X \gg 1$}).
\end{equation}
Thus we find that the tail is not the exponential nor the stretched Gaussian distributions, which are often observed in glass-forming liquids\cite{chaudhuri2007universal, saltzman2008large, kob1997dynamical, yamamoto1998heterogeneous} or some biological systems\cite{chechkin2017brownian,jeon2016protein, leptos2009dynamics, kurtuldu2011enhancement, he2016dynamic}.

\section{Relation to other systems}
The motion of molecule $A$ with $\mu\ll 1$ can be considered to be similar to that in the Lorenz gas model\cite{andersen1964relaxation}, which has been widely investigated as a dynamic model for light gas molecules in spatially fixed obstacles \cite{boldrighini1983boltzmann, dorfman2021,moran1987diffusion, machta1983diffusion}.
In the Lorentz gas model, only the velocity direction changes
and the speed remains unchanged at any timescale.
Thus, the mechanism that causes the fluctuating diffusivity observed in our system cannot be realized in Lorentz gas systems.

Our results suggest that fluctuating diffusivity emerges if the mass contrast is large: $\mu\ll 1$. To the best of our knowledge, there is no experimental report on the non-Gaussian behavior and fluctuating diffusivity in gas systems. However, we speculate that fluctuating diffusivity can be realized in experiments for binary gas mixtures. For instance, in the binary gas mixture of helium and radon\cite{hirst1939diffusion}, the mass ratio is $\mu \approx 0.018$.
For such a mass ratio, the non-Gaussian behavior originating from fluctuating diffusivity can emerge as shown in Fig.~\ref{fig_appendix_ngp} in Appendix. We expect that the non-Gaussian behavior will be observed if elaborated and precise measurements are performed.
Although the kinetics of gases\cite{chapman1990} may be considered as almost fully understood, our results imply that they are not yet understood.

\section{Conclusion}
In this study, we identified a novel origin of fluctuating diffusivity, which is neither environmental heterogeneity nor conformational degrees of freedom.
Fluctuating diffusivity emerges in simple binary gas mixtures with mass and fraction contrasts when the mass of the minor component molecule is sufficiently small in comparison to that of the major component.
We showed that fluctuating diffusivity originates from the timescale separation between the relaxation times of the velocity direction and the speed of the minor component molecule.
Our findings open a new modeling path for fluctuating diffusivity.
They will also shed light on the kinetic behavior of gas systems from a new aspect.
We hope that the predicted non-Gaussian behavior and fluctuating diffusivity will be experimentally observed in the future.

FN was supported by a Grant-in-Aid (KAKENHI) for JSPS Fellows (Grant No.~JP21J21725 from the Ministry of Education, Culture, Sports, Science and Technology (MEXT).
TU was supported by JST PRESTO Grant No.~JPMJPR1992 from the Japan Science and Technology Agency (JST).

\appendix
\renewcommand\thefigure{\thesection.\arabic{figure}}
\setcounter{figure}{0}

\section{Collision statistics}
\label{appendix_collision_statistics}

The kinetic Monte Carlo (KMC) method requires collision statistics as inputs.
In the present case, the probability density of a collision for molecule $A$ is required, which can be derived based on 
the gas kinetic theory\cite{resibois1977, dorfman2021, mazenko2008}.
We employ the following assumptions on our system:
\begin{enumerate}
 \item
       \label{assumption_markovian}
       The dynamics of molecule $A$ obeys a Markovian stochastic process.
 \item \label{assumption_collison_frequency}
       Molecule $B$ is homogeneously distributed in space.
\end{enumerate}
From assumptions~\ref{assumption_markovian} and \ref{assumption_collison_frequency}, the collision rate at which molecule $A$ with velocity $\bm{v}_{A}$ collides with molecule $B$ with velocity $\bm{v}_{B}$ can be expressed as follows:
\begin{equation}
 \label{collision_rate}
 \rho\sigma^{2}(\bm{v}_{A}-\bm{v}_{B})\cdot\hat{\bm{r}}_{AB} \Theta[(\bm{v}_A-\bm{v}_{B})\cdot\hat{\bm{r}}_{AB}], 
\end{equation}
where $\Theta(x)$ denotes the Heaviside step function
(no collision occurs when $(\bm{v}_{A}-\bm{v}_{B})\cdot \hat{\bm{r}}_{AB}< 0$).

The required probability density for collision $P(\bm{v}_{B}, \hat{\bm{r}}_{AB}, s|\bm{v}_{A})$ can be decomposed into the product of  three factors. The first factor is the cumulative waiting-time distribution of molecule $A$ with velocity $\bm{v}_{A}$.
Owing to the Markovian nature of the dynamics, this factor becomes an exponential distribution.
The second factor is the probability density of the velocity of the colliding molecule $B$, which is the Maxwell-Boltzmann velocity distribution.
The third factor is the collision rate, given by Eq.~\eqref{collision_rate}.
For the calculation of the first factor, the probability density of the waiting time $s$ is required:
\begin{equation}
 \label{waiting_time_distribution}
 P(s | \bm{v}_{A}) = F(\bm{v}_{A})e^{-F(\bm{v}_{A}) s}.
\end{equation}
Here, $F(\bm{v}_{A})$ is the average collision frequency and is expressed as follows:
\begin{equation}
\begin{split}
  F(\bm{v}_A) &= \int d\bm{v}_{B} d\hat{\bm{r}}_{AB} \,
\rho \sigma^{2} ( \bm{v}_{B}-\bm{v}_{A} ) \cdot \hat{\bm{r}}_{AB}
\Theta((\bm{v}_{B}-\bm{v}_A)\cdot\hat{\bm{r}}_{AB})
P_{\text{MB}}(\bm{v}_B ; m_B)\\
 &= \frac{\rho \pi \sigma^2}{\sqrt{\alpha}}
\left[\left(\sqrt{\alpha} |\bm{v}_{A}|+\frac{1}{2\sqrt{\alpha}|\bm{v}_{A}|}\right)\mathrm{erf}(\sqrt{\alpha} |\bm{v}_{A}|)
+\frac{1}{\sqrt{\pi}}\exp\left(-\alpha |\bm{v}_{A}|^2\right)\right],
\end{split}
 \label{eq_collision_frequency_f}
\end{equation}
where $\alpha=\beta m_{B}/2$.
The first factor is the probability of no collisions occurring during time $s$, which is calculated as follows \cite{Visco2008}:
\begin{equation}
 \Psi(s | \bm{v}_{A} ) = \int_{s}^{\infty} ds' \, P(s' | \bm{v}_{A}) = e^{-F(\bm{v}_{A}) s}.
\label{eq_prob_not_collide}
\end{equation}
The second factor is simply expressed as
\begin{equation}
 \label{maxwell_boltzmann_distribution_velocity}
 P_{\text{MB}}(\bm{v}_{B};m_{B}) = \left(\frac{\beta m_{B}}{2 \pi}\right)^{3/2}
  \exp\left( - \frac{\beta m_{B} \bm{v}_{B}^{2}}{2} \right).
\end{equation}
The probability density $P(\bm{v}_{B}, \hat{\bm{r}}_{AB}, s|\bm{v}_{A})$ can be expressed as
\begin{equation}
 \label{full_collsion_probability}
 P(\bm{v}_{B}, \hat{\bm{r}}_{AB}, s|\bm{v}_{A}) = 
   \Psi(s | \bm{v}_{A} )  P_{\text{MB}}(\bm{v}_{B};m_{B})
   \rho\sigma^{2}(\bm{v}_{A}-\bm{v}_{B})\cdot\hat{\bm{r}}_{AB} \Theta[(\bm{v}_A-\bm{v}_{B})\cdot\hat{\bm{r}}_{AB}].
\end{equation}
Equations~\eqref{full_collsion_probability}, ~\eqref{eq_prob_not_collide}, and \eqref{maxwell_boltzmann_distribution_velocity},
give Eq.~(3) in the main text.

\section{Numerical scheme for KMC simulation}
\label{appendix_scheme}

The collision-based dynamics of molecule $A$ can be
simulated using the KMC method\cite{gillespie1976general,bortz1975new} with Eq.~\eqref{full_collsion_probability} as the input.
The initial velocity of molecule $A$ is sampled 
based on Maxwell-Boltzmann distribution. The probability density of the initial velocity represented in dimensionless units is
\begin{equation}
 \label{maxwell_boltzmann_distribution_velocity_initial_state}
 P_{\text{MB}}(\bm{v}_{A};\mu) = \left(\frac{\mu}{2 \pi}\right)^{3/2}
  \exp\left( - \frac{\mu \bm{v}_{A}^{2}}{2} \right). 
\end{equation}
where $\mu$ is the mass ratio $m_A/m_B$, the same as in the main text.
Since Eq.~\eqref{maxwell_boltzmann_distribution_velocity_initial_state}
is a Gaussian distribution, $\bm{v}_A$ can be sampled using the Box-Muller method\cite{Devroye1986}.

For the time evolution of the system,
sampling of the stochastic variables $\bm{v}_{B}$,
$\hat{\bm{r}}_{AB},$ and $s$ are required.
However, the simultaneous sampling of these variables is technically difficult.
Therefore, we decompose the probability density 
$P(\bm{v}_B, \hat{\bm{r}}_{AB}, s | \bm{v}_A)$ into several conditional
probability densities as follows:
\begin{equation}
 P(\bm{v}_B, \hat{\bm{r}}_{AB}, s | \bm{v}_A)
= P(\hat{\bm{r}}_{AB} | \bm{v}_B, s, \bm{v}_A)P(\bm{v}_B | s, \bm{v}_{A})P(s| \bm{v}_{A}),
\end{equation}
where $P(\hat{\bm{r}}_{AB} | \bm{v}_B, s, \bm{v}_A)$, $P(\bm{v}_B | s, \bm{v}_A)$, and $P(s| \bm{v}_A)$ are defined as follows
\begin{equation}
 \label{eq_appendix_prob_tau} 
 P(s|\bm{v}_A)=\int d\bm{v}_B d\hat{\bm{r}}_{AB} \,
P(\bm{v}_B, \hat{\bm{r}}_{AB}, s | \bm{v}_{A}) 
=F(\bm{v}_A)e^{-F(\bm{v}_{A})s},
\end{equation}
\begin{equation}
 P(\bm{v}_B|s, \bm{v}_{A}) = \int d\hat{\bm{r}}_{AB}\, \frac{P(\bm{v}_B, \hat{\bm{r}}_{AB}, s | \bm{v}_{A})}{P(s|\bm{v}_{A})}
=\frac{\rho\pi\sigma^2|\bm{v}_{A}-\bm{v}_B|P_{\mathrm{MB}}(\bm{v}_B;m_{B})}{F(\bm{v}_A)}, \label{eq_appendix_prob_v} 
\end{equation}
\begin{equation}
 P(\hat{\bm{r}}_{AB} | \bm{v}_B, s, \bm{v}_{A})
=\frac{P(\bm{v}_B, \hat{\bm{r}}_{AB}, s | \bm{v}_{A})}{P(\bm{v}_B | s, \bm{v}_{A})P(s| \bm{v}_{A})}
=\frac{1}{\pi}\frac{\bm{v}_{A}-\bm{v}_B}{|\bm{v}_{A}-\bm{v}_B|}\cdot \hat{\bm{r}}_{AB} \Theta[(\bm{v}_{A}-\bm{v}_B)\cdot \hat{\bm{r}}_{AB} ]. \label{eq_appendix_prob_r}
\end{equation}
$F(\bm{v}_{A})$ in dimensionless units becomes 
\begin{equation}
  F(\bm{v}_A) = 
   \pi \left( |\bm{v}_{A}| + 1 / |\bm{v}_{A}| \right) \mathrm{erf}(|\bm{v}_{A}| / \sqrt{2})
+ \sqrt{2 \pi} \exp\left(-|\bm{v}_{A}|^2 / 2\right).
 \label{eq_collision_frequency_f_dimensionless}
\end{equation}
Based on these decomposed probability densities, $s$, $\bm{v}_{B}$, and $\hat{\bm{r}}_{AB}$ can be sampled
sequentially.
$s$ can be sampled using the inversion method\cite{Devroye1986} with
Eqs.~\eqref{eq_appendix_prob_tau} and \eqref{eq_collision_frequency_f_dimensionless}, respectively.

Equation~\eqref{eq_appendix_prob_v} can be rewritten with the relative velocity, $\bm{v}_{r} = \bm{v}_{B} - \bm{v}_{A}$.
Without loss of generality, the relative velocity can be expressed by spherical coordinates according to $\bm{v}_{r} = v_{r} \cos \phi \sin \theta \bm{e}_{x} + v_{r} \sin \phi \sin \theta \bm{e}_{y} + v_{r} \cos \theta \bm{e}_{z}$.
Here, $\bm{e}_{x}, \bm{e}_{y}$ and $\bm{e}_{z}$ are orthonormal basis vectors and $\bm{e}_{z}$ is set to $\bm{e}_{z} = \bm{v}_{A} / |\bm{v}_{A}|$.
Subsequently, Eq.~\eqref{eq_appendix_prob_v} is reduced to
\begin{equation}
 P(v_r, \theta, \phi|s, \bm{v}_{A})=
\frac{1}{4 (2\pi)^{3/2} F(\bm{v}_A) }
v_r^3\sin\theta
\exp\left[-(v_r^2/2+|\bm{v}_A|^2/2+|\bm{v}_A|v_r\cos\theta)\right].\label{eq_conditional_polar}
\end{equation}
Because $\phi$ is not included in Eq.~\eqref{eq_conditional_polar}, $\phi$ 
can be sampled from the uniform distribution.
The conditional probability density of $v_r$ is obtained by integrating Eq.~\eqref{eq_conditional_polar} over $\theta$ and $\phi$ as follows:
\begin{equation}
\label{eq_conditional_polar_vr}
\begin{split}
 P(v_r | s, \bm{v}_{A})
=&\int d\theta d\phi \, P(v_r, \theta, \phi | s, \bm{v}_{A}) \\
=&\left[\frac{\pi}{(2\pi)^{3/2} |\bm{v}_A| F(\bm{v}_{A})}
\exp\left(-\frac{|\bm{v}_A|^2}{2}\right)\right]  v_r^2\exp\left(-\frac{v_r^2}{2}\right)
\sinh\left(|\bm{v}_A|v_r\right).
\end{split}
\end{equation}
$v_{r}$ can be sampled using the rejection method\cite{Devroye1986} with Eq.~\eqref{eq_conditional_polar_vr}.
The conditional probability density of $\theta$ is:
\begin{equation}
\label{eq_conditional_polar_theta}
\begin{split}
  P(\theta | v_r, \phi, s, \bm{v}_{A}) 
&= \int d\phi \, \frac{P(v_r, \theta, \phi | s, \bm{v}_{A})}{P(v_r | \tau, \bm{v}_{A})}\\
&= \left[\frac{|\bm{v}_A|v_r}{2\sinh\left(|\bm{v}_A|v_r\right)}\right]
\sin\theta \exp\left(-v_r|\bm{v}_A|\cos\theta\right) .
\end{split}
\end{equation}
Subsequently, $\theta$ can be sampled using the inversion method.
$\bm{v}_{B}$ is obtained from sampled $v_{r}$, $\theta$, and $\phi$.

In a similar manner,
Eq.~\eqref{eq_appendix_prob_r} can be simplified using spherical coordinates.
Without loss of generality, $\hat{\bm{r}}_{AB}$ can be expressed as
$\hat{\bm{r}}_{AB} = \cos \phi' \sin \theta' \bm{e}_{x'} + \sin \phi' \sin \theta' \bm{e}_{y'} + \cos \theta' \bm{e}_{z'}$.
Here, $\bm{e}_{x'}$, $\bm{e}_{y'}$, and $\bm{e}_{z'}$ are orthogonal basis vectors and $\bm{e}_{z'}$ is set to $\bm{e}_{z'} = -\bm{v}_r/|\bm{v}_r|$.
Subsequently, Eq.~\eqref{eq_appendix_prob_r} can be expressed as:
\begin{equation}
 P( \theta', \phi' | \bm{v}_B, s, \bm{v}_{A})
=\frac{1}{\pi}\cos\theta' \sin\theta' \Theta(\cos\theta'). \label{eq_appendix_prob_theta_phi}
\end{equation}
Equation~\eqref{eq_appendix_prob_theta_phi} does not depend on $\phi'$. Therefore, $\phi'$ can be sampled from a uniform distribution, and
$\theta'$ can be sampled using the inversion method with
Eq.~\eqref{eq_appendix_prob_theta_phi}.
$\hat{\bm{r}}_{AB}$ can be constructed 
from $\theta'$ and $\phi'$.

\section{Additional Simulation Data}
\label{appendix_data}

\begin{figure}[ht!]
\begin{center}
  \includegraphics[width=80mm]{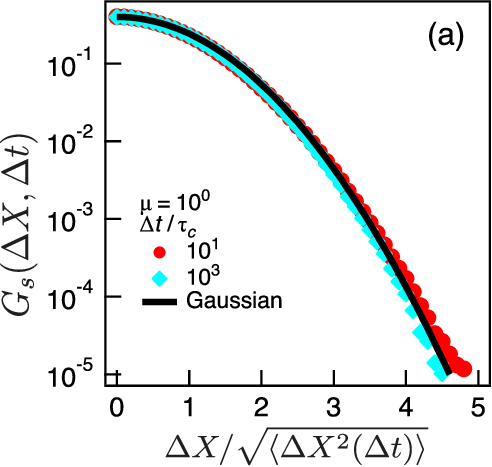}
\includegraphics[width=80mm]{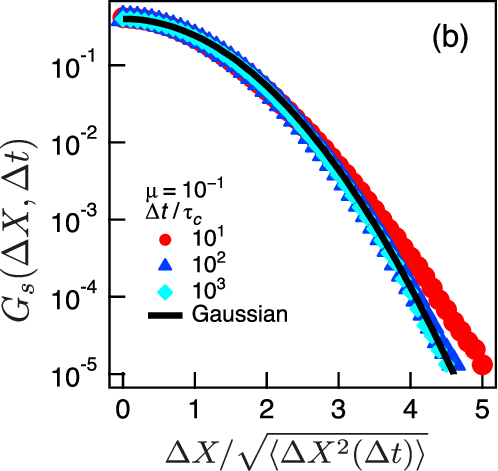}
\includegraphics[width=80mm]{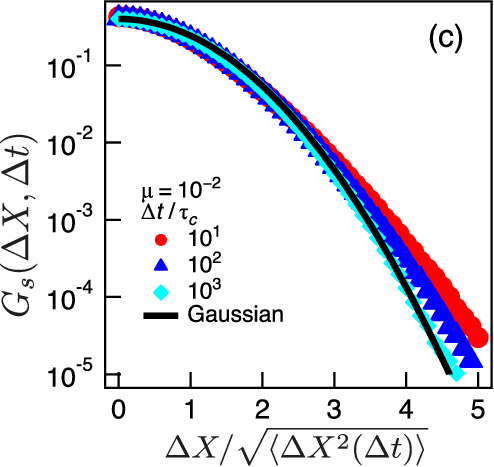}
\includegraphics[width=80mm]{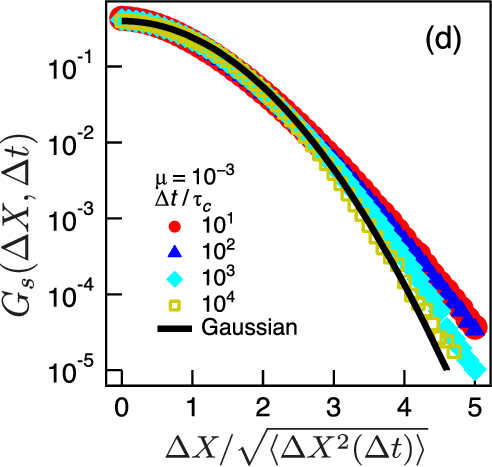}
\caption{The self-part of the van Hove correlation function of molecule $A$ for (a) $\mu= 10^{0}$, (b) $10^{-1}$, (c) $10^{-2}$, and (d) $10^{-3}$ from the KMC simulation. The displacement is normalized using the root mean square displacement $\sqrt{\langle \Delta X^2(\Delta t) \rangle}$. The solid curves represent the Gaussian distribution.}
\label{fig_appendix_van_hove}
\end{center}
\end{figure}

\begin{figure}[ht!]
\begin{center}
    \includegraphics[width=100mm]{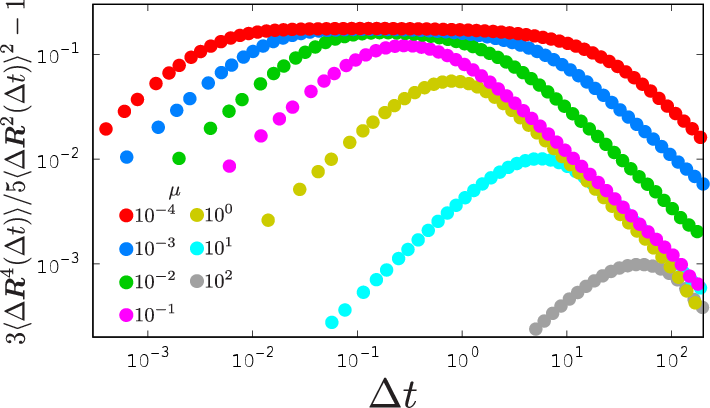}
    \caption{The non-Gaussian parameter of the molecule $A$ defined as 
    $3\langle \Delta\bm{R}^4(\Delta t) \rangle/5\langle \Delta\bm{R}^2(\Delta t) \rangle^2-1$ with various mass ratios from the KMC simulation.
    }
    \label{fig_appendix_ngp}
\end{center}
\end{figure}

\begin{figure}[ht!]
 \begin{center}
  \includegraphics[width=100mm]{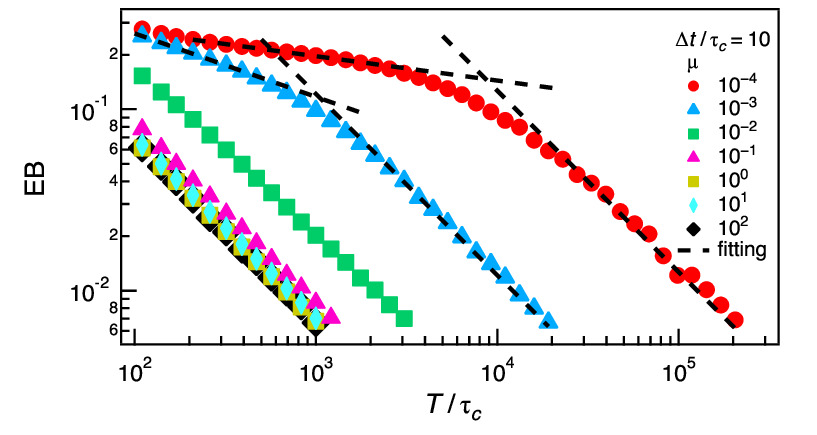}
\caption{The ergodicity breaking (EB) parameter of molecule $A$ for various $\mu$ from the KMC simulation. The dotted lines represent the curve fittings according to the power laws $\text{EB}\propto T^{-\alpha}$ and $\text{EB}\propto T^{-1}$ at the short and long-time regions.
The shoulder and the crossover behavior can be observed only for
$\mu=10^{-3}$ and $10^{-4}$.}
\label{fig_appendix_eb}
 \end{center}
\end{figure}

\begin{figure}[ht!]
 \begin{center}
  \includegraphics[width=80mm]{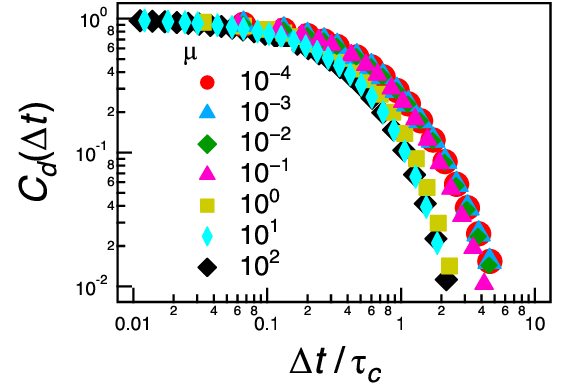}
\caption{Direction correlation function of molecule $A$, $C_d(\Delta t)$, for various mass ratios $\mu$, from the KMC simulation.}
\label{fig_appendix_cd}
 \end{center}
\end{figure}

\begin{figure}[ht!]
 \begin{center}
  \includegraphics[width=120mm]{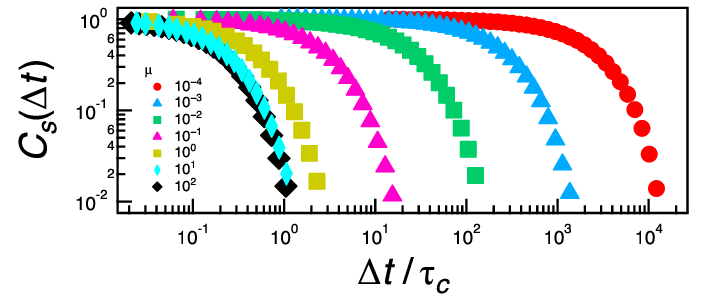}
\caption{The speed correlation function of the molecule $A$, $C_s(\Delta t)$,  for various mass ratios $\mu$, from the KMC simulation.}
\label{fig_appendix_cs}
 \end{center}
\end{figure}

\begin{figure}[ht!]
 \begin{center}
  \includegraphics[width=100mm]{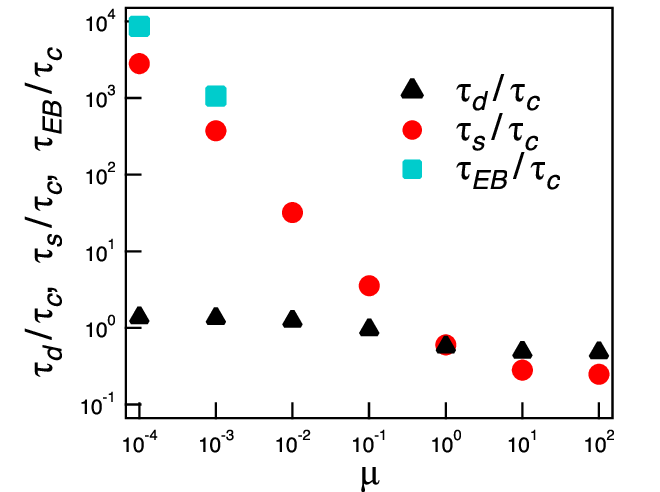}
\caption{The direction and speed relaxation times $\tau_{d}$ and $\tau_{s}$, and the crossover time $\tau_{\text{EB}}$ from the KMC simulation data in Figs.~\ref{fig_appendix_eb}-\ref{fig_appendix_cs}. $\tau_{\text{EB}}$ is estimated only for $\mu = 10^{3}$ and $10^{4}$.}
\label{fig_appendix_tau}
 \end{center}
\end{figure}

In the main text, we showed the representative simulation data only with mass ratios $\mu=10^{-4}$ and $10^2$.
In this Appendix, we show the results with different mass ratios $10^{-4}\le \mu \le 10^2$.
The self-part of the van-Hove correlation functions of the molecule $A$ with $\mu=$(a)$10^0$, (b)$10^{-1}$, (c)$10^{-2}$, and (d)$10^{-3}$ are displayed in Fig.~\ref{fig_appendix_van_hove}.
The non-Gaussian parameters against time lag with various $\mu$ are shown in Fig.~\ref{fig_appendix_ngp}.
Fig.~\ref{fig_appendix_eb} displays the EB parameters with various $\mu$. 
Figs.~\ref{fig_appendix_cd} and \ref{fig_appendix_cs} show the time-correlation functions of the direction and the speed of the molecule A.
From the data in Figs.~\ref{fig_appendix_eb}-\ref{fig_appendix_cs},
we estimate the characteristic timescales for EB, direction, and speed.
The characteristic timescale for EB can be estimated as the crossover time, as explained in the main text. The characteristic time scales for the direction and time are estimated as
\begin{equation}
 \tau_\gamma  = \frac{\displaystyle \int_{0}^{\infty} d\Delta t \, \Delta t C_\gamma(\Delta t)}{\displaystyle  \int_{0}^{\infty} d\Delta t \, C_\gamma(\Delta t)},
\end{equation}
with $\gamma = d, s$. These estimates are displayed in Fig.~\ref{fig_appendix_tau}.

\bibliographystyle{apsrev4-1}
\bibliography{ref.bib}
\end{document}